\documentclass[conference]{IEEEtran}
\IEEEoverridecommandlockouts

\usepackage[T1]{fontenc}
\usepackage[hyphens]{url}

\usepackage[a4paper,top=2cm,bottom=30mm,left=3cm,right=3cm,marginparwidth=1.75cm]{geometry}

\usepackage{amsmath}
\usepackage{graphicx}

\usepackage{cite}
\usepackage{amsmath,amssymb,amsfonts}
\usepackage{algorithmic}
\usepackage{graphicx}
\usepackage{textcomp}
\usepackage{xcolor}
\def\BibTeX{{\rm B\kern-.05em{\sc i\kern-.025em b}\kern-.08em
    T\kern-.1667em\lower.7ex\hbox{E}\kern-.125emX}}
\begin{document}

\title{Towards a Cloud Fog Edge System for Smart Buildings
}


\author{\IEEEauthorblockN{Christophe Cérin}
\IEEEauthorblockA{\textit{University Sorbonne Paris Nord} \\
\textit{LIPN, UMR CNRS 7030}\\
\textit{INRIA DATAMOVE}\\
Paris/Grenoble, France \\ 
christophe.cerin@univ-paris13.fr}
\and
\IEEEauthorblockN{Mamadou Sow}
\IEEEauthorblockA{\textit{University Sorbonne Paris Nord} \\
\textit{LIPN, UMR CNRS 7030} \\
Villetaneuse, France \\
mamadou.sow@univ-paris13.fr}
\and
\IEEEauthorblockN{Frédéric Andrès}
\IEEEauthorblockA{\textit{National institute of Informatics} \\
2 Chome-1-2 Hitotsubashi,\\
Chiyoda City,\\ 
Tokyo 101-8430, Japan\\
andres@nii.ac.jp}}

\maketitle

\begin{abstract}
In this article, we present our vision and recent advancements toward creating a decentralized system capable of learning from real-time data within buildings to support sustainable and privacy-preserving smart environments. Our approach promotes the concept of the building itself as the data center, aligning with the principles of edge computing to safeguard confidentiality and reduce reliance on external cloud infrastructure. This is particularly valuable in humanitarian contexts, where data sovereignty, energy efficiency, and infrastructure constraints are critical. We detail a lightweight, "Kubernetes-like" orchestration framework for deploying AI services within such environments and demonstrate our progress in implementing AI algorithms on low-power, cost-effective microcontrollers such as those in the Arduino ecosystem. By enabling in-situ learning directly on sensors or microcontrollers, our work aims to bring intelligent services to resource-limited settings, fostering autonomy, resilience, and sustainable development in vulnerable or underserved communities. The contributions in this article are related, firstly, to our project "Online Machine Learning Algorithms for Embedded Systems" and the evaluation of two new online algorithms. Secondly, we envision a cloud-fog-edge architecture based on the KOptim and FIWARE components, and we propose a methodology for coupling them. Experimental results of the online algorithms are also presented, showcasing real-world traces.
\end{abstract}


\section{Introduction}\label{sec:one}


The principal aim of this paper is to advance inclusive and sustainable technology research in the domains of smart building Information and Communication Technologies (ICT), electrical engineering, communications, and edge computing. With a humanitarian focus, we explore technological solutions that minimize environmental impact and are accessible to under-resourced or marginalized communities. Our approach advocates for alternatives to energy-intensive cloud computing and centralized infrastructures, proposing instead ecologically responsible, locally deployable systems based on cost-effective microcontrollers.

The ICT sector is increasingly scrutinized for its substantial environmental footprint, driven by high energy consumption in computation, storage, and networking infrastructures~\cite{iea-AI-Energy-2025,AI-generative-X24}. In 2023, global electricity consumption from ICT was estimated at around 2\,\%, with the figure rising to 4.4\,\% in the United States~\cite{lbl}. Forecasts suggest that global Data Center (DC) electricity usage may reach 3\% by 2030~\cite{iea2025}. In addition to carbon emissions, artificial intelligence (AI) and high-performance computing (HPC) technologies require significant water and rare metal resources, the latter often being non-recyclable. These trends raise critical questions about the sustainability and equity of digital transformation, particularly in humanitarian contexts where resource constraints are pronounced.

Simultaneously, electricity use in buildings accounted for nearly 60\% of global consumption growth in 2024, with natural gas demand also rising by 1\,\% in residential and commercial sectors~\cite{iea2025}. The intersection of energy-intensive computing and increasing building energy use presents a compelling case for rethinking infrastructure. This paper promotes edge computing as a potential pathway to mitigate these pressures, emphasizing in-situ intelligence that reduces the need for data transmission and centralized processing. While the overall environmental cost of edge computing relative to cloud services remains underexplored, the promise of reduced energy usage, enhanced data privacy, and local autonomy warrants deeper investigation.

Our research problem is thus framed by the needs of individuals, public institutions, or private entities managing activity within buildings—whether optimizing operations through embedded AI or regulating high-usage environments to conserve resources. We present two use cases aligned with these goals. In both, we explore the design and deployment of an embedded AI system built on edge computing principles. Such a system must operate independently within the building, continuously learning from real-time sensor data to support resilient, efficient, and sustainable outcomes.


The structure of this paper is organized as follows: In Section \ref{sec:two}, we provide definitions relevant to our domain, present a usage scenario, and review related literature. Section \ref{sec:trois} details two categories of experiments undertaken, encompassing the materials employed and the methodologies adopted. The initial experiment focuses on cloud orchestration in conjunction with FIREWARE, while the subsequent experiments pertain to the application of online machine learning algorithms at the edge. Section \ref{sec:four} offers an analysis and interpretation of the results, situating them within the established theoretical framework. We also explain the drawbacks of our proposed method. Finally, Section \ref{sec:five} offers conclusions and proposes directions for future research.

\section{Definitions and Related Works}\label{sec:two}

\subsection{Important definitions}

NIST, a US government agency, defines cloud computing as "{\it a model for on-demand network access to a shared pool of configurable resources (e.g., networks, servers, storage, applications, and services) that can be quickly provisioned and released with minimal management or provider effort}". This IT shift offers many benefits, redefining the landscape. Cloud computing's flexibility surpasses on-site solutions, allowing data and application access from anywhere with the Internet. Its scalability, accessibility, security, and collaborative features make it vital for enterprises and research centers, promoting innovation, agility, and digital transformation.

Edge computing executes programs close to the request sources for rapid response. In his IEEE DAC 2014 keynote \cite{DBLP:conf/vts/Arabi21}, Karim Arabi defined it as computing outside the cloud, at the network edge, for applications requiring fast data processing.

NIST defines IoT (Internet of Things \cite{iot-computing}) as "{\it the network of devices with the hardware, software, firmware, and actuators to connect and exchange data.}" IoT connects devices to the Internet and is related to edge computing, though they differ: edge computing processes data locally, while IoT devices typically send data to the cloud. As IoT devices gain computing power\footnote{\url{https://www.st.com/en/microcontrollers-microprocessors/stm32n6-series.html}}, this distinction is fading. Fog computing's importance is growing, with nodes closer to users reducing network latencies. Fog computing \cite{fog-computing} enables efficient data processing via intermediary nodes between the cloud and IoT objects.

\begin{figure}
    \centering
    \includegraphics[width=0.95\linewidth]{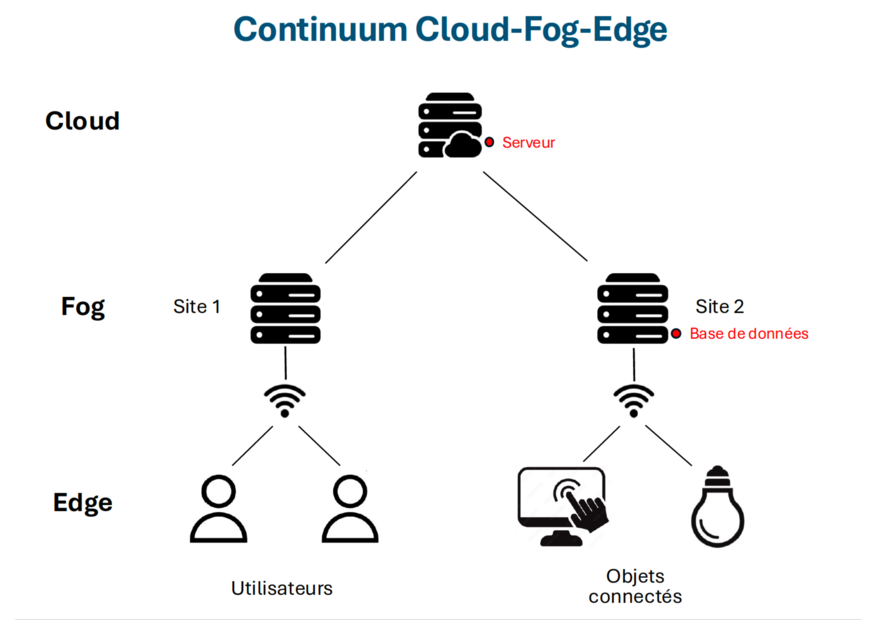}
    \caption{Architectural concerns}
    \label{fig:one}
\end{figure}

Figure \ref{fig:one} illustrates that we aim to build a system that balances architectural designs across cloud, fog, and edge nodes. Let us imagine that a municipality needs to manage its social housing stock and its district heating network using a digital tool. This municipality has a private cloud (see top of Figure \ref{fig:one}) that communicates with the fog level, which corresponds to servers installed in each social housing building. The fog level communicates with the edge level, which corresponds to individual housing units where temperature, CO2, and other sensors are installed. In this architectural framework, we need to deploy, for example, AI algorithms on fog nodes to learn building temperature and occupancy, enabling the prediction of temperature or occupancy changes in parts of the building. In the latter case, this could involve learning the intensity score, as defined by M2intenses\footnote{\url{https://intensiscore.m2intenses.com/}}. Indeed, buildings and spaces, whether public or private, are largely underutilized. Yet, better exploitation would offer potentially colossal ecological, economic, and social benefits. How can we maximize the use of buildings?

%
%

\subsection{Related work on cloud orchestrator}

Contemporary cloud orchestrator frameworks predominantly depend on container management tools engineered by diverse corporations and open source communities, including Docker SwarmKit~\cite{SwarmKit}, Apache Mesos~\cite{mesosapache}, Red Hat OpenShift~\cite{OpenShift}, and Google Kubernetes~\cite{GoogleKubernetes}.

Docker SwarmKit~\cite{SwarmKit} is crucial for container scheduling, assigning containers to nodes using filtering and selection. Mesos~\cite{HBKAZMGAJADKRHSSSI11}, an Apache project, is a resource-sharing layer for cluster frameworks, with Marathon~\footnote{\url{https://mesosphere.github.io/marathon/}} managing applications on it. Marathon launches containers anywhere in the Mesos cluster, ensuring fault tolerance, but it can also place them on specific nodes using constraints.

RedHat OpenShift~\cite{OpenShift} is a container orchestration platform enabling web application development and deployment without dedicated servers. Managed via CLI or Web interface~\cite{LAPARWA17}, it supports backend services, APIs, stateful apps, big data, AI/ML, edge, serverless, and event-driven computing.

Google Kubernetes~\cite{GoogleKubernetes} is a leading open-source tool for automating the deployment, scaling, and management of container-based applications. It employs Pods, which can operate on-premises or in public clouds. Clusters consist of a Kubernetes master and several worker nodes. The scheduler assigns Pods to nodes by first filtering out nodes that don't meet the requirements, leaving Pods pending if resources are lacking, and then scoring and selecting the best node from the remaining options.

In this work, we highlight the latest advances in KOptim \cite{DBLP:conf/ipps/MenouerDCR25}, which offers a generic framework for scheduling and resource management. Compared to current Kubernetes-based frameworks and scheduling strategies, KOptim has two main benefits: (i) it simplifies container resource configuration through Service Level Agreement (SLA) classes, and (ii) it enhances online container scheduling using a Multi-Criteria Decision Making (MCDM) algorithm. KOptim serves as a generic framework for various Kubernetes distributions, including K3s and MicroK8s, which are tailored to Fog and Edge computing environments. The crucial reason for using KOptim is its ability to manage dynamic SLA, meaning that instead of specifying strict constraints on CPU, RAM, networking, and disk resources, we rely on a range of values, thus providing more flexibility and adapting to resource-constrained devices. However, in this paper, we do not explain all the workings of the KOptim mechanics due to space constraints in the article, and we refer the reader to the bibliography.

\subsection{Related work on AI for smart buildings}

Building operating costs are mainly due to HVAC systems, which require monitoring of indoor temperature, weather, equipment consumption, and occupancy. Secondary factors enhancing thermal understanding include precise schedules, sunshine, and HVAC heat/power ratio. Environmental and equipment data are critical for smart buildings, integrating through BIM (Building Information Management) and BMS (Building Management Systems). 

We consider online machine learning algorithms \cite{wang2024comprehensivesurveycontinuallearning, hoi2018onlinelearningcomprehensivesurvey} where data arrive sequentially and update the predictor at each step. Batch learning generates a predictor by processing all training data at once. Continuous learning involves ongoing model improvement amid continuous data, handling the gradual influx of new information. The approach involves utilizing an offline algorithm constrained by a window of limited size, which is particularly necessary given the limitations inherent in embedded devices. We implement drift control through various methodologies; for example, by organizing the data post-processing by the offline algorithm, followed by sorting and systematically selecting and eliminating data points at regular intervals. This method aims to maintain diversity throughout the process.

\subsection{Related work on IoT benchmarking, workload emulation characterization}

The work done for this article has its roots in the project "Online Machine Learning for Embedded Systems"\footnote{https://github.com/christophe-cerin/OnlineML\_ESP32}, whose objective is to offer online algorithms for embedded hardware. The project is a collection of algorithms in Python or C/C++ covering the most useful machine learning algorithmic expressions.

This project is closely related to complementary needs in our context, as those in \cite{huang:hal-04382496,Batz2021IoTWE,DBLP:conf/edge2/TadakamallaM19}. In \cite{huang:hal-04382496}, the authors introduce U-TOE, a universal toolkit intended to support IoT designers and researchers. It integrates features of a low-power embedded operating system, a generic model transpiler and compiler, a built-in performance evaluation module, and an open-access remote IoT testbed. The authors release an open-source implementation of U-TOE and illustrate how it can be leveraged to experimentally assess the performance of different neural network models on a broad spectrum of low-power IoT boards built on widely used microcontroller architectures. U-TOE enables straightforward, reproducible, and customizable comparative performance studies across diverse IoT hardware platforms simultaneously. In contrast, our work concentrates not on neural network methods but on online machine learning algorithms.

In \cite{Batz2021IoTWE}, the authors present a straightforward yet broadly applicable method to model the workload of representative IoT applications, aiming to provide a realistic and reproducible means of emulating IT loads for data centers. Their method decomposes IoT applications into application primitives, which are then used to emulate the workload that the modeled application would generate. The paper ends with a case study demonstrating IoT Workload Emulation in the BodenTypeDC experiment, where innovative cooling strategies for data centers were evaluated under realistic workload conditions. By contrast, in our scenario, the architecture is fixed through the use of KOptim and FIWARE.

In \cite{DBLP:conf/edge2/TadakamallaM19}, the authors tackle the problem of workload characterization, which is a fundamental step for conducting performance and Quality of Service engineering analyses. A system’s workload is defined as the collection of all inputs it receives from its environment over one or more time intervals. Characterizing this workload involves identifying the nature of its basic components and providing a quantitative and probabilistic description of these components in terms of arrival processes, event counts, and service demands. Although multiple workload characterization studies have been proposed across various domains, IoT workloads have largely been overlooked. This gap is precisely what the paper aims to fill, and it additionally includes a capacity planning study grounded in one of the workload characterizations it introduces. By contrast, we do not perform workload characterization; instead, we directly employ real-world traces (rather than synthetic data) to assess our online algorithms. In other words, the paper does not provide a characterization of the traces it uses.

\section{Experiments}\label{sec:trois}

\subsection{Materials and methods}


Our experimental research is organized into two distinct branches. The first branch focuses on the implementation of the proposed cloud orchestrator integrated with a FIWARE component \cite{SLADashboard,SLAFramework}. The FIWARE Foundation is instrumental in the establishment and implementation of open-source standards crucial to the development of modular and interoperable smart solutions. These standards facilitate the creation of such solutions efficiently, economically, and quickly, while mitigating the risk of vendor lock-in and fostering FIWARE as a sustainable and innovation-centric business ecosystem. In our case, we propose a FIWARE standard to make the cloud orchestrator interoperable with a database of policies for dynamic Service Level Agreements.

The second branch pertains to the implementation of machine learning algorithms tailored for the Arduino ecosystem, specifically adapted for the microcontroller context. We illustrate the challenges of building online solutions, and we conduct experiments on real datasets from the smart building sector.



\subsection{Results}


\subsubsection{KOptim and FIWARE from a methodological point of view}


In paper \cite{KOptim_2024}, KOptim is introduced as a Kubernetes framework devised under the fogSLAs project \cite{https://doi.org/10.1002/spe.70058}, aiming to streamline resource management by associating Service Level Agreements (SLAs) with containers in a cloud infrastructure. Despite frequent discussions around SLAs, the deployment or sharing of open-source solutions remains infrequent. The novel contribution of KOptim lies in its reliance on dynamic SLAs, signifying that these agreements can evolve. To achieve this, a range of values has been implemented in place of the conventional fixed resource quantities typical of traditional cloud environments.

Current open-source Service Level Agreement (SLA) implementations predominantly concentrate on defining structures rather than on providing comprehensive toolkits for production management. Notwithstanding, they exhibit a standard pattern for SLA parameters, necessitating the definition of a range to indicate compliance and a timeframe to evaluate any violations. Among the array of open-source solutions, it is observed that FIWARE \cite{SLADashboard,SLAFramework} uniquely offers extensive management tools, including an SLA definition framework and a dashboard for monitoring active SLAs. Considering FIWARE's aspiration to establish itself as a standard in distributed systems, it presents an ideal platform for the development of our dynamic SLA prototype. The fogSLA project holds the potential to augment Fiware's capacity to support pure microservice or \emph{as-a-service} hosting methodologies.

\begin{figure}[hbtp]
    \centering
    \includegraphics[width=0.95\linewidth]{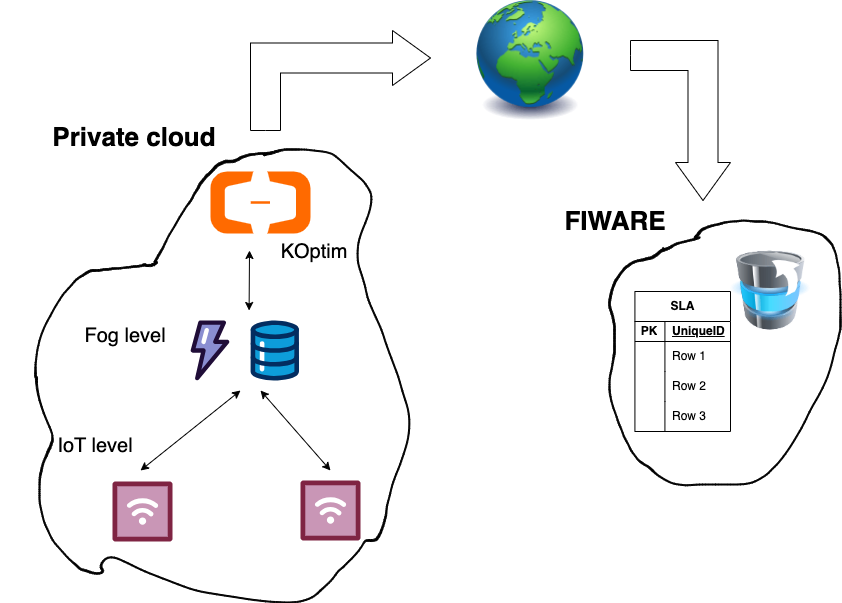}
    \caption{KOptim coupled with FIWARE}
    \label{koptim-arch}
\end{figure}

We implemented an architecture in which the KOptim scheduler communicates with the FIWARE environment. We used the WS-Agreement scripting from FIWARE's SLA Framework to define SLA classes for managing KOptim container limits. We are now working to set up a monitoring solution for dynamic resources and SLOs. The complete architecture is depicted in Figure \ref{koptim-arch}, where we assume that an operator in the Smart Building sector manages a private cloud based on KOptim. The KOptim scheduler deploys containers for running online machine learning algorithms (see forthcoming subsections). Koptim can deploy services on microcontrollers, but also on fog nodes (see Figure \ref{fig:one}) to collect information produced at the edge level. Depending on the traffic between the fog and edge levels, some fog nodes can become busy, necessitating a new negotiation of the resources associated with services. At this time, KOptim asks the FIWARE component to find the "best" policy and resource number that does not violate the SLA expressed by the smart building operator.

It should be noted that, as depicted in Figure \ref{koptim-arch}, the FIWARE components are proposed to reside outside the private cloud under the administration of the building operator. This configuration replicates our actual conditions since the experiment’s source code, and consequently, the service code source pertinent to scheduling, remains inaccessible to the public owing to constraints imposed by the KOptim framework.

The results we now detail are those for which the codes are available online\footnote{\url{https://github.com/christophe-cerin/OnlineML_ESP32/tree/main/ARDUINO/}}.

\subsubsection{Data mining from the Perret Tower with the GHA dimension reduction algorithm}


This study focuses on applying a dimensionality reduction algorithm to a dataset originating from the Tour Perret building in Grenoble. Dimensionality reduction involves the transformation of data from a high-dimensional space into a lower-dimensional space, ensuring that the reduced representation retains essential characteristics of the original data, ideally approximating its intrinsic dimension. The Tour Perret dataset comprises up to 15 dimensions. Our objective is to achieve a two-dimensional representation to facilitate more effective visual analysis. In this context, we demonstrate the implementation of the GHA-PCA algorithm\cite{SANGER1989459}.

We now discuss the adaptation of the abstract algorithm to fit our requirements for running on embedded devices and, more importantly, to transform it into an online algorithm.

The developed code utilizes the Eigen library for matrix operations, which is a popular choice in C++ for linear algebra tasks. This C++ code retains the core functionality of the original Python code while adapting it to the syntax and conventions of C++. The program will process the data in blocks of 1024 lines of the Perret Tower data ~\cite{TourPerrethead10col.csv}, display the iteration number at each step, and update the eigenvalues and eigenvectors accordingly. In the end, the results will be saved in a CSV file and displayed graphically.
 
\paragraph{Explanation} We have introduced a constant W that defines the size of the buffer (1024 lines). The main loop now runs through the data in blocks of 1024 lines. At each iteration, we display the current iteration number and the lines currently being processed. For each block of 1024 lines, we apply the GHA algorithm to each line of the block. If the total number of lines is not a multiple of 1024, the last block will be smaller. We use $\min(W, n-i)$ to handle this. The Eigen template library already manages memory dynamically.

The header file ghapca.h contains the declarations of variables, constants, and shared functions needed for the source code\footnote{\url{https://github.com/madou-sow/OnlineML_ESP32/blob/main/ARDUINO/GHA-PCA/src/online_GhaPca_update_buffer.cpp}} to be compiled into a file. In a second program\footnote{\url{https://github.com/madou-sow/OnlineML_ESP32/blob/main/ARDUINO/GHA-PCA/src/online_GhaPca_update_buffer_normalize.cpp}}, there is a special feature, namely, a section where the data is normalized. This practice allows for transforming the data without distorting it. Normalization consists of harmonizing the data so that all the entries of the different data sets that relate to the same terms follow a similar format.
    
\paragraph{Analysis} The differences between the results in C++ and Python\footnote{\url{https://github.com/madou-sow/OnlineML_ESP32/blob/main/MICROPYTHON-ULAB/DimensionalityReduction/testing_batch_ghapca_json.py}} programs can be attributed to several factors. The C++ program processes the data in blocks of 1024 rows, while the Python program processes the data sequentially. This may affect how the eigenvalues and eigenvectors are updated. The initial parameters, such as the eigenvalues and the eigenvectors, may be initialized differently in the two programs, which may influence the final results. The C++ program uses an online approach to update the eigenvalues and eigenvectors, while the Python program uses a batch approach. This may lead to differences in the convergence of the results. The Python program ignores rows without payload, which can reduce the number of data rows processed, while the C++ program does not seem to have this logic. The C++ program normalizes the data by ignoring some columns (1, 3, and 6), while the Python program normalizes all columns. This can lead to differences in the values of the processed data. The possible causes are i) Constant column: Columns 1, 3, and 6 may contain a constant value (e.g., all elements are 0, 1, or some other fixed value).; ii) Missing or poorly formatted data: If columns 1,3, and 6 contain missing or poorly formatted data, it could lead to a situation where all values are the same after processing; iii) Data loading issue: The CSV file might not be loaded properly, resulting in incorrect values in columns 1, 3, and 6.
    
\paragraph{How to bypass the previous difficulties related to data quality} We implemented the following solutions. If columns 1, 3, and 6 are not important for your analysis, you can ignore them when normalizing. If columns 1, 3, and 6 are important but contain constant values, you can replace these values with a default value or a small variation to allow normalization. If normalization is a problem, you can consider using standardization (subtract the mean and divide by the standard deviation) instead of normalization. This works even if the values are constant (although the standard deviation is zero in this case, which would also require special handling).

\subsubsection{Online machine learning and Online technique with CampusIOT's LoRaWAN Datasets and the K-means Clustering Algorithm}

Again, in this subsection, our goal is to design a model that can learn from new data without revisiting past data, as well as one that is robust to concept drift. The concept of drift refers to the evolutionary changes in the data that render an existing data model obsolete. This phenomenon occurs when there are unforeseen alterations in the statistical properties of the target variable that the model is designed to predict over time. The data set used also comes from the Smart Building ecosystem. This developed model will be very close to what happens in a production context, which is generally event-based. It will integrate well with the rest of the embedded systems programming ecosystem.

In the simulation, the CO2 concentration and temperature are published from our local MQTT server with the subjects "CO2/Celsius" and "final/final". The data is received via the ESP32 board and is used by the Arduino IDE, where a program~\cite{esp32MqttKmeansClusterQuickBuffer} has been implemented to perform clustering with the K-MEANS algorithm. In our case, we considered a structure as a means of defining this type of data, although it is more than a simple container for primitive data types because we can also define other functionalities. What it does with the elements or data points is a movement that meets the criteria at the end of the range. It then returns an iterator to the first element of the deleted elements (which are just moved).

The objective is to analyze the behavior of the ESP32 controller in the face of a stream of messages received continuously, subject to an implemented algorithm containing both a clustering algorithm, which makes it possible to identify a maximum of 4 clusters.

During the experiments, we managed to send several message streams (256, 512, 1024, and 5120) containing CO2 quantities and temperatures from the MQTT server to the ESP32 microcontroller. Using our integrated Arduino development environment, we have carried out a distribution of the data to be processed by the microcontroller. This variable is called W. It varies from 8 to 512. 512 is the maximum value, taking into account the saturation that can occur.

By referring to an equation with 3 unknowns, namely x (set of messages), y (splitting or subset of messages), and z (number of splits of messages processed), the satisfactory result corresponds to the case that allows the observation of the maximum number of messages processed in a message set.

\begin{figure}[hbtp]
    \centering
    \includegraphics[width=0.95\linewidth]{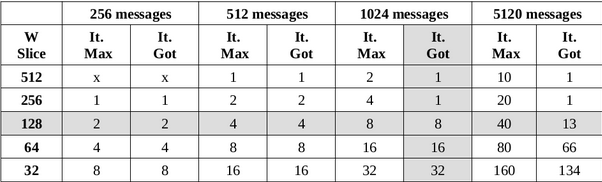}
    \caption{Message Flow, Processing Slice and Number of Possible Iterations}
    \label{fig:three}
\end{figure}

Looking at the table in Figure \ref{fig:three}, we can say that the ESP32 microcontroller can easily receive a stream of 1024 messages with optimized batch processing of 128 messages.

\subsubsection{KMeans Clustering for the CampusIOT's LoRaWAN Datasets}

We consider here the "ELSYS-ERS-CO2, ERS-CO2-48B30 W=512, Data = 5120, K=4" use case, which is the settings. This input file contains the basic information (ID, CO2, T(°C), Cluster) about the data recorded. Clustering involves dividing points in a data set into several groups such that data points in the same group are more similar to other data points in the same group than to those in other groups. The goal in clustering is to separate groups with similar characteristics and distribute them into clusters.

\begin{figure}[htbp]
    \centering
    \includegraphics[width=0.95\linewidth]{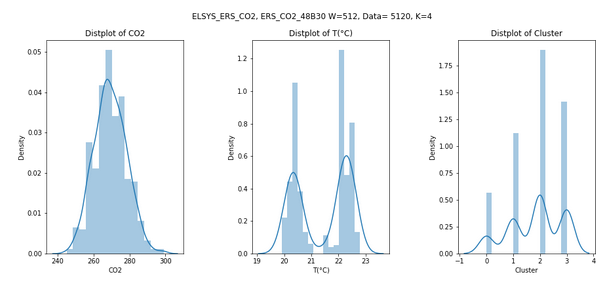}
    \caption{Distribution Diagram ELSYS\_ERS\_CO2, ERS\_CO2\_48B30 W=512, Data= 5120, K=4}
    \label{fig:four}
\end{figure}

We define a target number k, which refers to the number of centroids we need in the dataset. As for the centroid, it is the imaginary or real location representing the center of the cluster. Each point in the dataset is assigned to each of the clusters by reducing the sum of squares within the cluster. Alternatively, the K-means algorithm identifies k number of centroids and then allocates each data point to the closest cluster, while keeping the centroids as small as possible. The “means” in K-means refers to find the center of gravity of the data. Using the Seaborn module as well as the Matplotlib module, we will represent the distribution diagram with different variations of the data, such as cluster1data5120w512.csv.  A distribution diagram in the Figure \ref{fig:four} represents the variation in the distribution of data by a histogram and a line in combination with it. 

\section{Discussion}\label{sec:four}


This work aims to develop an AI-based building management system with two main components. The first automates AI algorithm deployment to manage load fluctuations by integrating dynamic SLAs. We propose using the established open-source FIWARE standard to decouple SLA management. This is necessary as the Kubernetes-based scheduler remains a closed system, accessible through FIWARE.

We have developed and tested two online AI algorithm examples that are foundational to machine learning and useful in the context of smart buildings. The key challenge is to create memory-efficient solutions for our microcontroller library. Our solutions apply to the ARDUINO ecosystem, and some algorithms are also designed for the Micropython ecosystem, another platform for microcontroller programming.

Regarding the development aspect of our work, limitations include a time lag between ecosystem algorithm developments (ARDUINO and Micropython), an incomplete understanding of online algorithms for building management, and the need for feedback on additional datasets.

More generally, it's important to revisit the assumptions underlying this work, which some may disagree with. For example, we assumed that the cloud orchestrator was ultimately a black box that we didn't control. Therefore, the orchestrator is assumed to run in a private cloud that we don't own.

On the other hand, we assumed that the SLA management module, using FIWARE, which is open source (for the part we're interested in), could be deployed within the building, along with the measurement sensors. The "building is the data center" approach is, in our case, somewhat loose. We could very well use a cloud orchestrator that would also be deployed within the building. Our choice is driven by the fact that a building manager (a municipal authority, for example) often manages several buildings, so deploying an orchestrator for each building is undoubtedly more expensive. This is a sizing issue that, while important, is not addressed in our article.

Furthermore, on the downside, we believe our proposal is less suitable for a situation where we need to train a general model (in the machine learning sense) for all buildings. Indeed, this scenario would require uploading all the data to the cloud, which can generate network traffic. However, by placing an orchestrator in each building and using federated learning techniques \cite{10.1561/2200000083}, we could simply exchange model "summaries" to build a general model available in each building. This could be a potential extension of our work.

\section{Conclusion and future works}\label{sec:five}

This paper seeks to consolidate clearly formulated perspectives for edge-AI researchers, system architects, policymakers, and domain specialists, enabling them to investigate approaches, frameworks, and governance structures that uphold responsible autonomy and transparency in practical AI deployments. 


This paper outlines our perspective and recent progress towards developing a decentralized system that learns from real-time building data to support smart and sustainable environments with a focus on privacy. We propose treating the building itself as the data hub, adhering to edge computing concepts to protect privacy and lessen dependence on external cloud services. 

This is especially important in humanitarian settings, where data ownership, energy conservation, and infrastructure limitations are crucial. We describe a lightweight orchestration framework akin to "Kubernetes" for the deployment of AI services within these environments and exhibit our advances in applying AI algorithms on low-power, affordable microcontrollers, such as those in the Arduino ecosystem. By facilitating on-site learning directly on sensors or microcontrollers, our initiative aims to introduce intelligent services to resource-constrained settings, enhancing autonomy, resilience, and sustainable growth in vulnerable or underserved communities.

In summary, our future work aims to address the challenges we introduced in the discussion section, specifically those related to the delay in advances in ecosystem algorithms (ARDUINO and Micropython), a better understanding of online AI algorithms for building management, the need for feedback on additional datasets, and finally, those related to federated learning if the building manager needs to have a general model of what is happening in the buildings they manage.


\bibliographystyle{IEEEtran}
\bibliography{IEEEabrv,mybib}

\end{document}